\documentclass[aps,11pt,nofootinbib,preprintnumbers,groupedaddress,superscriptaddress]{revtex4-2}
\usepackage{amssymb, amsfonts, amsmath, bm}
\usepackage[
]{graphicx}
\usepackage{color}
\usepackage[
]{hyperref}
\hypersetup{%
 setpagesize=false,
 bookmarksnumbered=true,%
 bookmarksopen=true,%
 colorlinks=true,%
 linkcolor=blue,%
 citecolor=red}

\begin{document}
\title{
Periapsis shifts in dark matter distribution with a dense core
}
\author{Takahisa Igata}
\email{takahisa.igata@gakushuin.ac.jp}
\affiliation{Department of Physics, 
Gakushuin University, Mejiro, Toshima-ku, Tokyo 171-8588, Japan}
\affiliation{KEK Theory Center, 
Institute of Particle and Nuclear Studies, 
High Energy Accelerator Research Organization, Tsukuba 305-0801, Japan}
\author{Yohsuke Takamori}
\email{takamori@wakayama-nct.ac.jp}
\affiliation{National Institute of Technology (KOSEN), 
Wakayama College, Gobo, Wakayama 644-0023, Japan}
\date{\today}
\preprint{KEK-TH-2395, KEK-Cosmo-0285}

\begin{abstract}
We consider the periapsis shifts in dark matter distribution with a dense core. 
We model the dark matter distribution as 
an isotropic gas sphere, 
the Emden polytropic sphere of index 5 in general relativity.
This model has a parameter range where all the energy conditions are satisfied in the entire region. 
Within the parameter range, the asymptotic analysis 
for stellar 
motion allows us to identify two competing effects on the bounded motion: the general-relativistic effect and a local-density effect of 
matter.
Furthermore, using nearly circular bound orbits, we demonstrate that
retrograde periapsis shifts occur near the center, where the local-density effect dominates over the general-relativistic effect, 
whereas prograde periapsis shifts occur in the far region, 
where the general-relativistic effect dominates over the local-density effect.
This result means that a natural explanation for the retrograde periapsis shifts is not the existence of exotic objects (e.g., naked singularities or wormholes) but the local distribution of physically reasonable matter on the stellar orbit. Furthermore, it also implies that the periapsis shift plays a crucial role in distinguishing black hole alternatives, such as dark matter cores, from a pure black hole. 
\end{abstract}
\maketitle

\section{Introduction}
\label{sec:1}
Sagittarius A* (Sgr A$^\ast$) is a massive and compact radio source located at the center of our Galaxy. 
This source is surrounded by a cluster of stars in close orbits~\cite{Ghez:2008ms,Genzel:2010zy}. 
The observations of individual stars (S-stars) demonstrate that each of the dynamics is well described by a test particle approximation in the gravitational field of Sgr A$^\ast$~\cite{Abuter:2018drb,Do:2019txf,Saida:2019mcz,Abuter:2020dou,Takamori:2020ntj}.
This suggests that S-stars are ideal probes for studying the nature of the central object as a gravitational source and its surroundings.

One of the main concerns about Sgr A* is the true identity---whether it is a black hole or an alternative such as a naked singularity, a wormhole, and an exotic compact object. 
Answering this question requires further investigations of the central region, not only the visible components 
but invisible mass distribution.
The observations of S2/S0-2, the closest S-star to the center, have reported so far that upper limits on the total dark mass inside the orbit ($\lesssim 0.01\ \mathrm{pc}$) are less than 
1\% of the Sgr A*~\cite{
Do:2019txf,Abuter:2020dou}.

In fact, it was pointed out that the effect of the dark component around Sgr A$^\ast$ can be observed through the periapsis shift phenomenon of S-stars. 
In a vacuum spacetime centered on a black hole, the prograde periapsis shifts of stellar bound orbits occur due to the general-relativistic effect (see, e.g., Ref.~\cite{Weinberg:1972}). 
In contrast, post-Newtonian analysis reveals that the effect of matter distribution decreases the periapsis shift angle and even cancels out the prograde shift due to the general-relativistic effect~\cite{Rubilar:2001}. 
This result suggests that even a dark component with sufficiently small mass relative to the central object can still affect the dynamics of the surrounding stars.
For example, suppose that the distribution has a region of locally high energy density, through which a star passes. Then the Ricci curvature, which takes a large value there, affects the stellar dynamics through the metric. 
However, it remains unclear how matter distribution contributes to the retrograde shift in the fully general-relativistic regime.

Recently, it has been pointed out that the retrograde periapsis shifts can occur in spacetimes centered on naked singularities~\cite{Bambhaniya:2021ybs,Ota:2021mub}.
This suggests that we can use the periapsis shift phenomenon in distinguishing between black holes and their alternatives. 
However, the presence of naked singularities often creates a physically unacceptable situation in which associated matter violates energy conditions.

In view of these situations, it is useful to consider the periapsis shift phenomenon in a black hole alternative spacetime, which consists only of physically reasonable dark matter with a dense core region. Therefore,
the purpose of this study is to model the distribution of dark matter with a dense core as a solution to the Einstein equations and to clarify the mechanism for determining the sign of the periapsis shift angles of stellar bound orbits in that spacetime. 
Thus, this study focuses on a general-relativistic generalization of the Newtonian Plummer model, described by the Buchdahl spacetime~\cite{Buchdahl:1964}.
It may be useful for a qualitative understanding of the competing effects of general relativity and local matter density distribution because the solution is characterized by two parameters and has a physically reasonable parameter range in the sense of the energy conditions.

This paper is organized as follows. 
In Sec.~\ref{sec:2}, we briefly review 
the Buchdahl spacetime.
In Sec.~\ref{sec:3}, we formulate the dynamics of a freely falling stellar object 
in this spacetime. Furthermore, we discuss two competing effects of the general-relativistic correction and the local matter density distribution on the bounded motion.
In Sec.~\ref{sec:4}, we consider the nearly circular 
bound orbits of stars and 
clarify the appearance of prograde and retrograde periapsis shifts. 
Section~\ref{sec:5} is devoted to a summary and discussion. 
Throughout this paper, we use geometrized units in which $G=1$ and $c=1$.

\section{Buchdahl spacetime}
\label{sec:2}
We review the Buchdahl spacetime~\cite{Buchdahl:1964}.
The metric is given by
\begin{align}
\label{eq:metric}
\mathrm{d}s^2
&=-\frac{(1-f)^2}{(1+f)^2}\:\!\mathrm{d}t^2
+(1+f)^4 (\mathrm{d}r^2+r^2\:\!\mathrm{d}\theta^2+r^2\sin^2\theta \:\!\mathrm{d}\varphi^2),
\\
f(r)&=\frac{a}{2\sqrt{1+k r^2}},
\end{align}
where $a$ and $k$ are constants. 
The physical interpretation of these parameters becomes clear in Eqs.~\eqref{eq:Mdef} and \eqref{eq:Rdef} below.
The spatial part is conformally isometric to the Euclidean flat metric and is written by the standard spherical coordinates~$(r, \theta, \varphi)$. 
This metric admits stationarity and spherical symmetry. 
We assume 
that $a>0$ and $k>0$. 
Then, the function $f$ takes the value $f(0)=a/2>0$ at the center $r=0$ and 
decreases monotonically with $r$ to zero, where $0\le r< \infty$.

The Einstein equations lead to matter distribution being a perfect fluid with a stress-energy tensor,
\begin{align}
T_{ab}=\rho \:\!u_a u_b+p\:\!(g_{ab}+u_a u_b),
\end{align}
where $g_{ab}$ denotes the metric tensor, $u^a$ denotes the tangent field of static observers filling the spacetime, and 
\begin{align}
\label{eq:rho}
\rho(r)&=\frac{24kf^5}{\pi a^4(1+f)^5},
\\
\label{eq:p}
p(r)&=\frac{8kf^6}{\pi a^4(1-f^2)(1+f)^4}
\end{align}
are energy density and pressure, respectively. 
We also introduce $q$ as 
\begin{align}
q=\frac{3p}{\rho}=\frac{f}{1-f}.
\end{align}
From Eqs.~\eqref{eq:rho} and \eqref{eq:p}, we obtain the equation of state of the fluid,
\begin{align}
\label{eq:eos}
\frac{p}{p_{\mathrm{c}}}
=\frac{(\rho/\rho_{\mathrm{c}})^{6/5}}{1+2 q_{\mathrm{c}}\left[\:\!
1-(\rho/\rho_{\mathrm{c}})^{1/5}
\:\!\right]},
\end{align}
where $\rho_{\mathrm{c}}$, $p_{\mathrm{c}}$, and $q_{\mathrm{c}}$ are 
the values of $\rho$, $p$, and $q$ evaluated at the center $r=0$, respectively, 
\begin{align}
\rho_{\mathrm{c}}
&=\frac{24 ak}{\pi (2+a)^5},
\\
p_{\mathrm{c}}
&=\frac{8a^2 k}{\pi(2-a)(2+a)^5},
\\
q_{\mathrm{c}}
&=\frac{a}{2-a}.
\end{align}
If $q_{\mathrm{c}}\ll 1$ (i.e., $a\ll 1$), the equation of state~\eqref{eq:eos} reduces to the polytropic one with 
the polytropic index $5$, and the gravitational potential, the energy density, and the pressure of the Newtonian Plummer model are recovered.%
\footnote{
\begin{align}
\Phi&=-2f=-\frac{a}{\sqrt{1+k r^2}}.
\quad 
\rho
=\frac{3ka}{4\pi \left(1+kr^2\right)^{5/2}},
\quad 
p=\frac{a^2 k}{8\pi (1+kr^2)^3},
\end{align}
where $\Phi$ corresponds to the Newtonian gravitational potential. 
}
Therefore, this model is known as the 
general-relativistic 
Plummer model.

Imposing energy conditions on the matter field further restricts the parameter region of the solution. Several energy conditions are written as follows: (i) weak energy condition, $\rho\geq 0$ and $\rho+p\geq 0$; (ii) strong energy condition, $\rho+3p\geq 0$ and $\rho+p\geq 0$; (iii) null energy condition, $\rho+p\geq 0$; and (iv) dominant energy condition, $\rho\geq |p|$. 
For $a\leq 2$, $\rho$ and $p$ are non-negative in the entire region, thus satisfying the weak, strong, null energy condition. 
In contrast,
the dominant energy condition does not hold at least near the center for $3/2<a\leq 2$ but holds in the entire region for $a\leq 3/2$. 
Therefore, we assume $a\leq 3/2$, thus satisfying all the energy conditions in the entire region.

\begin{figure}[t]
\centering
\includegraphics[width=6.3cm,clip]{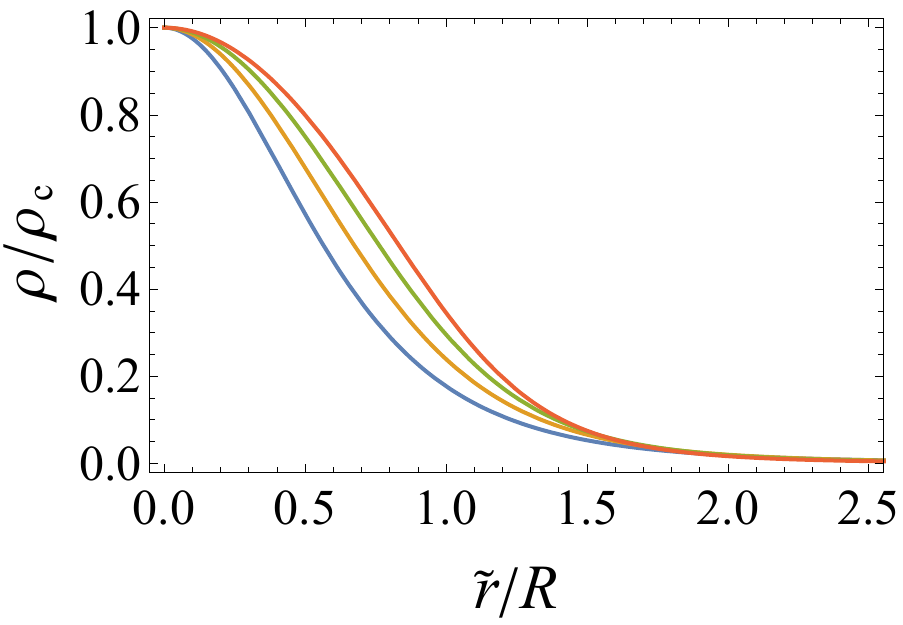}
\includegraphics[width=8.8cm,clip]{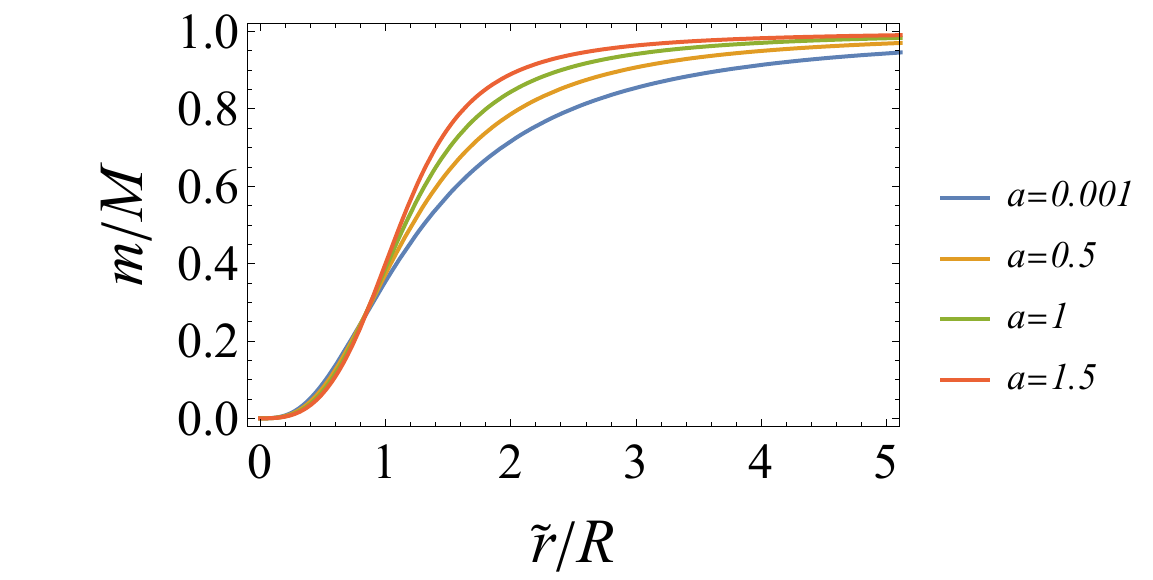}
\caption{Energy density distribution $\rho$ and the proper mass function~$m$. 
Left: each $\rho$ normalized by the central value $\rho_{\mathrm{c}}$ is denoted as a function of the areal radius $\tilde{r}$ normalized by the core radius $R$. 
Right: each mass $m$ normalized by the total proper mass $M$ is denoted as a function of $\tilde{r}/R$. }
\label{fig:rhomass}
\end{figure}

We define the proper mass inside the radius $r$ by
\begin{align}
m(r)
&=\int_0^r \rho(\xi) \sqrt{h}\:\!\mathrm{d}^3x
=\frac{a kr^3}{(1+kr^2)^{3/2}}+\frac{3a^2}{16\sqrt{k}}\left[\:\!
\arctan (\sqrt{k} r)-\frac{\sqrt{k}r (1-kr^2)}{(1+kr^2)^2}
\:\!\right],
\end{align}
where $\sqrt{h}\:\!\mathrm{d}^3x=\xi^2[1+f(\xi)]^6 \sin \theta \:\!\mathrm{d}\xi\mathrm{d}\theta\:\!\mathrm{d}\varphi$ is the three-dimensional volume element 
on a static hypersurface. 
Thus, the total proper mass is given by
\begin{align}
\label{eq:Mdef}
M
&=\lim_{r\to \infty}m(r)
=\frac{a}{\sqrt{k}}\left(1+\frac{3\pi a}{32}\right). 
\end{align}
We now introduce the areal radius $\tilde{r}$, which has a much clearer physical interpretation than $r$, as 
\begin{align}
\tilde{r}=r(1+f)^2.
\end{align}
We define the core radius $R$ as $\tilde{r}$ evaluated at 
the core scale $r=1/\sqrt{k}$,
\begin{align}
\label{eq:Rdef}
R
&=\left.\tilde{r}\right|_{r=1/\sqrt{k}}
=\frac{1}{\sqrt{k}}\left(1+\frac{a}{2\sqrt{2}}\right)^2.
\end{align}
We call the region where $\tilde{r}\leq R$ the core
and call the ratio $M/R$ the compactness of the core. 
Since this ratio depends monotonically on $a$ and is reduced to $a$ for $a\ll 1$, 
we will somewhat loosely refer to $a$ itself as the compactness. 
Figure~\ref{fig:rhomass} shows $\rho/\rho_{\mathrm{c}}$ and $m/M$ as functions of the normalized areal radius $\tilde{r}/R$.
These figures imply that the mass distribution of this model is localized near the center, particularly inside the core radius $R$. 
Even if the compactness is small (i.e., $a\ll 1$), 
the quantities $\rho$ and $m$ at the core radius 
still take the values
\begin{align}
\left.\frac{\rho}{\rho_{\mathrm{c}}}\right|_{r=1/\sqrt{k}}
&=
\left(\frac{2+a}{2\sqrt{2}+a}\right)^5
\simeq \frac{1}{4\sqrt{2}}=0.1767\ldots,
\\
\left.\frac{m}{M}\right|_{r=1/\sqrt{k}}
&\simeq \frac{1}{2\sqrt{2}}=0.3535\ldots,
\end{align}
respectively. 
As the system becomes more compact, that is, as the compactness increases, 
these values increase accordingly.

\section{Formulation of the stellar dynamics}
\label{sec:3}
We consider stellar dynamics in 
the Buchdahl spacetime.
Assume that the local interactions of matter with the stellar object is negligible. 
Hence, the matter distribution contributes to the dynamics only through the gravitational field so that the motion is freely falling (i.e., geodesic motion). 
Let $p_a$ denote the canonical momentum of a freely falling test particle with unit mass. 
Then, the Hamiltonian $H$ is given by
\begin{align}
H
=\frac{1}{2}g^{ab}p_ap_b
=\frac{1}{2(1+f)^4}\left[\:\!
-\frac{(1+f)^6}{(1-f)^2}\:\!p_t^2+ p_r^2+\frac{1}{r^2} \left(
p_\theta^2+\frac{p_\varphi^2}{\sin^2\theta}
\right)
\:\!\right],
\end{align}
where $g^{ab}$ is the inverse metric. 
Since the metric~\eqref{eq:metric} admits a stationary Killing vector $\partial/\partial t$ and an axial Killing vector $\partial/\partial \varphi$, a test particle has constants of motion associated with these symmetries,
\begin{align}
E&=-p_a (\partial/\partial t)^a=-p_t,
\\
L&=p_a (\partial/\partial \varphi)_a =p_\varphi,
\end{align}
which are interpreted as energy and angular momentum, respectively. 
Without loss of generality, we may assume that the particle motion is 
in any case confined on the equatorial plane
$\theta=\pi/2$ because of the spherical symmetry, and therefore $p_\theta=0$. 

We focus on the constraint equation
\begin{align}
g^{ab} p_a p_b=-1.
\end{align}
Using the Hamilton equation $\dot{r}=\partial H/\partial p_r=p_r/(1+f)^4$, 
this condition is rewritten as 
\begin{align}
\label{eq:radial}
&\frac{\dot{r}^2}{2}+V=0,
\\
&V(r)=\frac{1}{2(1+f)^4}+\frac{L^2}{2r^2(1+f)^8}-\frac{E^2}{2(1-f^2)^2},
\end{align}
where the dot denotes differentiation with respect to the affine parameter. 
Now we derive the asymptotic form of this equation and 
investigate the role of each order of the effective potential. 
To compare the asymptotic form with the conventional expression, we rewrite Eq.~\eqref{eq:radial} in terms of the areal radius $\tilde{r}$,
\begin{align}
&\frac{\dot{\tilde{r}}^2}{2}+\tilde{V}=0,
\\
&\tilde{V}=\frac{\beta^2}{2}\left[\:\!
-\frac{E^2}{(1-f^2)^2}+\frac{1}{(1+f)^4} \left(\frac{L^2}{\tilde{r}^2}+1\right)
\:\!\right],
\end{align}
where $r=r(\tilde{r})$, and $\beta$ is the coordinate transformation factor
\begin{align}
\beta
=\frac{\mathrm{d}\tilde{r}}{\mathrm{d} r}
=(1+f)\left[\:\!1+\frac{a}{2} \frac{1-k r^2}{(1+kr^2)^{3/2}}\:\!\right].
\end{align}
Expanding $\tilde{V}$ by a sufficiently small compactness $a$ (i.e., $a\ll1$), we obtain an expression to linear order in $a$,
\begin{align}
\tilde{V}
=\left[\:\!
1-\frac{2a k \tilde{r}^2}{(1+k \tilde{r}^2)^{3/2}}
\:\!\right]\left(\frac{L^2}{2\tilde{r}^2}+\frac{1}{2}\right)
-\left[\:\!
1+\frac{2a}{(1+k \tilde{r}^2)^{3/2}}
\:\!\right]\frac{E^2}{2}+O(a^2).
\end{align}
Furthermore, up to linear order in $a$, the asymptotic expansion of $\tilde{V}$ yields 
\begin{align}
\tilde{V}
\simeq \frac{1-E^2}{2}-\frac{M}{\tilde{r}}
+\frac{L^2}{2\tilde{r}^2}
-\frac{M L^2}{\tilde{r}^3}
+\left(\frac{3}{2}-E^2\right) \frac{MR^2}{\tilde{r}^3},
\end{align}
where we have used the leading-order expressions 
$M= a/\sqrt{k}$ and $R=1/\sqrt{k}$ of Eqs.~\eqref{eq:Mdef} and \eqref{eq:Rdef}, respectively. 
The first term is a constant, giving the difference in energy from unit rest mass.
The following three terms have the same form as the conventional 
effective potential of particle motion around a point source of mass $M$ in general relativity (i.e., the Schwarzschild case). 
The last term---not appearing in the conventional expression---depends on the extended distribution of matter via the core radius $R$. 
This term has the same power $\tilde{r}^{-3}$ as the fourth general-relativistic correction term but with the opposite sign at least in the bounded motion (i.e., $E\leq 1$). 
In the series of circular orbits and nearly circular bound orbits, the angular momentum decreases with decreasing orbital radius, and hence the general-relativistic correction term dominates over the last term at far region, while the last term dominates over the general-relativistic correction term near the center. 
This implies that the effect of the extended matter distribution 
may compensate for the phenomena associated with the familiar general-relativistic correction term. 
One of the phenomena where the effects of extended matter distributions would be observed is the periapsis shift.

\section{Periapsis shifts in the Buchdahl spacetime}
\label{sec:4}
\begin{figure}[t]
\centering
\includegraphics[width=10cm,clip]{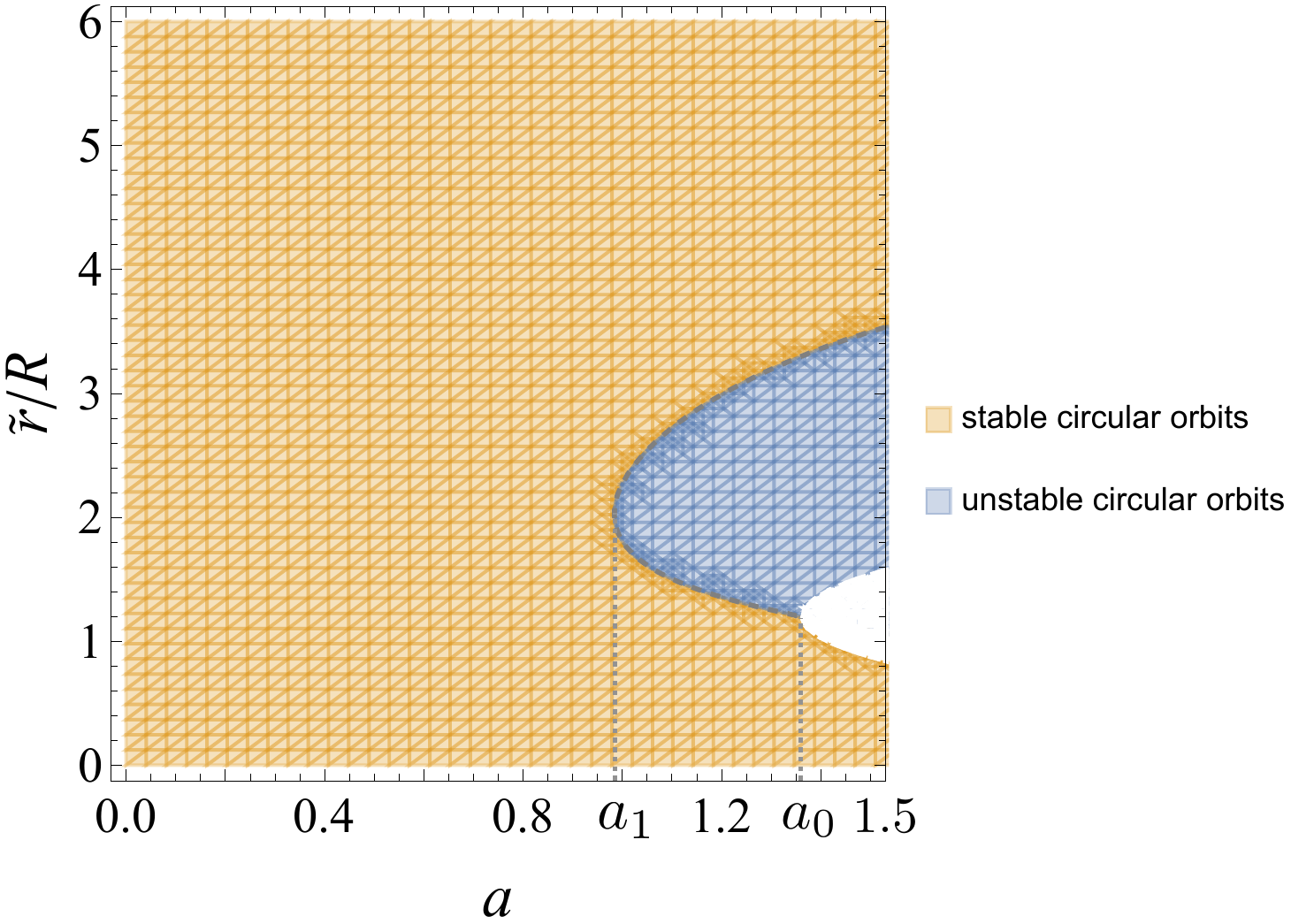}
\caption{Orbital radii of stable and unstable circular orbits of a star in the Buchdahl spacetime. The boundary
between these two shaded regions denotes marginally (un)stable circular orbits. 
The upper curve ($r=r^+_{\mathrm{ms}}$) and the lower curve ($r=r^-_{\mathrm{ms}}$) coincide with each other at $a=a_1$. 
There are no circular orbits in the uncolored region.
Its upper boundary ($r=r_{\mathrm{ph}}^+$) and lower boundary ($r=r_{\mathrm{ph}}^-$) show photon circular orbit radii, which coincide with each other at $a=a_0$.}
\label{fig:circular}
\end{figure}

We consider the periapsis shift phenomenon of a star in the Buchdahl spacetime. 
The equation of radial motion takes the form 
\begin{align}
\label{eq:ddot}
\ddot{r}=-V',
\end{align}
where the prime denotes differentiation with respect to $r$. 
As seen from this equation together with Eq.~\eqref{eq:radial}, 
the conditions for a star staying in a circular orbit, $\dot{r}=0$ and $\ddot{r}=0$, are given by
\begin{align}
V&=0, \\ 
V'&=0. 
\end{align}
Solving these equations for $E^2$ and $L^2$, we obtain 
\begin{align}
\label{eq:Er}
E^2(r)&
=\frac{(1-f)^3(1+f+2r f')}{(1+f)^2\left[\:\!
2\:\!r f'+(1-f)(1+f+2\:\!r f')
\:\!\right]},
\\
\label{eq:Lr}
L^2(r)&
=\frac{-2\:\! r^3 f' (1+f)^4}{2\:\!r f'+(1-f)(1+f+2\:\!rf')},
\end{align}
where $r$ denotes the orbital radius of a circular orbit. 
These quantities must take non-negative values for circular orbits. 
The circular orbits are stable if $V''> 0$, 
marginally (un)stable if $V''=0$,
and unstable if $V''< 0$, 
where the explicit form of $V''$ is given by
\begin{align}
\label{eq:V''}
V''
=\frac{3(1+f+r f')-2\:\!r^2 f''}{r^4 (1+f)^9}L^2(r)
-\frac{6f'^2(3-f)+2f'' (1-f^2)}{(1-f^2)^4} E^2(r).
\end{align}
Figure~\ref{fig:circular} shows the ranges of the circular orbit radii for various values of $a$. 
The orange and blue shaded regions show the regions where stable and unstable circular orbits exist, respectively. 
The boundary between the orange and blue regions denotes marginally (un)stable circular orbits. There are no circular orbits in the uncolored region.
For $0<a\le a_1$, stable circular orbits exist in the whole range of $r$, 
where 
\begin{align}
a_1=0.9842\ldots
\end{align}
is determined as the value of $a$ at the multiple root of $V''=0$. 
In contrast,
unstable circular orbits, one of the characteristic phenomena in the general-relativistic strong gravitational field, do not appear. This is interpreted as the compactness being so small that the effect of the matter distribution is strong enough to suppress the higher-order general-relativistic effects.
For $a_1\leq a<a_0$,%
\footnote{
The value $a_0$ is a positive real-valued solution to the equation
\begin{align}
3a^6-1512 a^4+10240a^2-13824=0.
\end{align}
The exact expression of $a_0$ is given by
\begin{align}
a_0=\left\{
168+\frac{16\sqrt{3809}}{3}
\cos\left[\:\!
\frac{1}{3}\arccos\left(
\frac{470097}{7618\sqrt{3809}}
\right)+\frac{2\pi}{3}
\:\!\right]
\right\}^{1/2}.
\end{align}
}
where 
\begin{align}
a_0=1.3629\ldots,
\end{align}
stable circular orbits exist only in $0\leq r\leq r^{-}_{\mathrm{ms}}$ or $r\ge r^+_{\mathrm{ms}}$, where $r^\pm_{\mathrm{ms}}$ are the orbital radii of the marginally stable circular orbits, determined by solving $V''=0$.
In contrast,
unstable circular orbits appear in 
$r^-_{\mathrm{ms}}\leq r\leq r^+_{\mathrm{ms}}$. 
The appearance of unstable circular orbits is a sign of the general-relativistic effects associated with increasing compactness.
The orbital radii $r^\pm_{\mathrm{ms}}$ coincide with each other at $a=a_1$, where $\sqrt{k} r^\pm_{\mathrm{ms}}=2.6788\ldots$ (i.e., $\tilde{r}_{\mathrm{ms}}^\pm/R=2.0254\ldots$). 
Note that the appearance of stable circular orbits near the center is a universal phenomenon when a regular center exists. In such cases, the effective potential is generally reduced to the two-dimensional isotropic harmonic oscillator type approximately. Correspondingly, in the current case, the expansion of $V$ around $r=0$ becomes
\begin{align}
\label{eq:V0}
V&=c_0+
\frac{128 L^2}{(2+a)^8 r^2}
+\frac{16 ak}{(2+a)^5} \left[\:\!
1+\frac{a(2+a)^2E^2}{(2-a)^3}+\frac{48(a-1) kL^2}{(2+a)^5}
\:\!\right]r^2+O(r^4),
\\
c_0&
=\frac{8}{(2+a)^4}\left[\:\!
1-\frac{(2+a)^2E^2}{(2-a)^2}+\frac{64 a k L^2}{(2+a)^5}
\:\!\right].
\end{align}
The second and third terms in Eq.~\eqref{eq:V0} balance to form the potential bottom, and hence a stable circular orbit appears there. 
For $a_0\leq a\leq 3/2$, stable circular orbits exist in 
$r\ge r^+_{\mathrm{ms}}$ or 
$0\le r< r_{\mathrm{ph}}^-$, 
and unstable circular orbits exist in 
$r_{\mathrm{ph}}^+<r<r^+_{\mathrm{ms}}$, 
where 
$r_{\mathrm{ph}}^\pm$
correspond to the radii of photon circular orbits, and 
$L(r)$ and $E(r)$ diverge in the limit $r\to r_{\mathrm{ph}}^\pm$.
The orbital radii 
$r_{\mathrm{ph}}^\pm$
coincide with each other at $a=a_0$, 
where $\sqrt{k}r_{\mathrm{ph}}^-=\sqrt{k}r_{\mathrm{ph}}^+= 1.3194\ldots$ 
(i.e., $\tilde{r}_{\mathrm{ph}}^-/R=\tilde{r}_{\mathrm{ph}}^+/R=1.1972\ldots$).
Note that there are no circular orbits in 
$r_{\mathrm{ph}}^-\leq r\le r_{\mathrm{ph}}^+$. 
The compactness is even larger, and a region near the center shows the absence of circular orbits due to the effect of strong gravity in general relativity. 
However, for the same reasons mentioned above, there are again stable circular orbits in the vicinity of the center.

If a star is displaced slightly from the ``equilibrium" radius $r$ of a stable circular orbit,
the particle will oscillate in radius about $r$.
Since the amplitude of the oscillation is sufficiently small compared to the equilibrium radius $r$, this orbit is nearly circular.
The frequency of the radial oscillation is given by 
\begin{align}
\omega_r=\sqrt{V''},
\end{align} 
where $V''$ is given by Eq.~\eqref{eq:V''}.
The angular frequency $\omega_\varphi$ is given by
\begin{align}
\omega_\varphi=\dot{\varphi}=\frac{L(r)}{r^2(1+f)^4}.
\end{align}
Then, we can introduce the precession rate for nearly circular bound orbits 
\begin{align}
\nu=\frac{\omega_\varphi-\omega_r}{\omega_\varphi}.
\end{align}
If $\nu>0$, then the prograde periapsis shift of the nearly circular bound orbit occurs; 
if $\nu=0$, the orbit is elliptic;
and if $\nu<0$, the retrograde periapsis shift occurs. 
Up to linear order in $a$, the asymptotic form of $\nu$ is reduced to%
\begin{align}
\label{eq:nuasym}
\nu\simeq \frac{3}{2} \left(
\frac{2M}{\tilde{r}}-
\frac{\rho}{\bar{\rho}}
\right),
\end{align}
where 
\begin{align}
\frac{\rho}{\bar{\rho}}\simeq \frac{R^2}{\tilde{r}^2}
\end{align}
and $\bar{\rho}=3m/(4\pi \tilde{r}^3)$ is the averaged energy density inside radius $\tilde{r}$.
The first term, proportional to $2M/\tilde{r}$, shows how close $\tilde{r}$ to the gravitational radius $2M$, indicating the general-relativistic effect. In contrast, 
the second term, proportional to $\rho/\bar{\rho}$, shows the magnitude of the matter 
local density, indicating the local-density effect. 
The asymptotic formula~\eqref{eq:nuasym} explicitly demonstrates that 
$\nu>0$ (i.e., the prograde periapsis shift) if the general-relativistic effect dominates over the local-density effect, whereas $\nu<0$ (i.e., the retrograde periapsis shift) if the local-density effect dominates over the general-relativistic effect.%
\footnote{The observable quantity for the periapsis shift is the shift angle $\Delta \varphi_{\mathrm{p}}$ per unit period of radial oscillation, 
\begin{align}
\Delta \varphi_{\mathrm{p}}
=2\pi \frac{\omega_\varphi-\omega_r}{\omega_r}
=2\pi\frac{\nu}{1-\nu}.
\end{align}
Up to linear order in $a$, the asymptotic form of $\Delta \varphi_{\mathrm{p}}$ is reduced to 
\begin{align}
\label{eq:Dphip}
\Delta \varphi_{\mathrm{p}}
\simeq 3\pi \left(\frac{2M}{\tilde{r}}-\frac{\rho}{\bar{\rho}}\right). 
\end{align}
The first term is the same as the conventional general-relativistic term, whereas the second term is a contribution from the local-density effect of matter.}

\begin{figure}[t]
\centering
\includegraphics[width=16cm,clip]{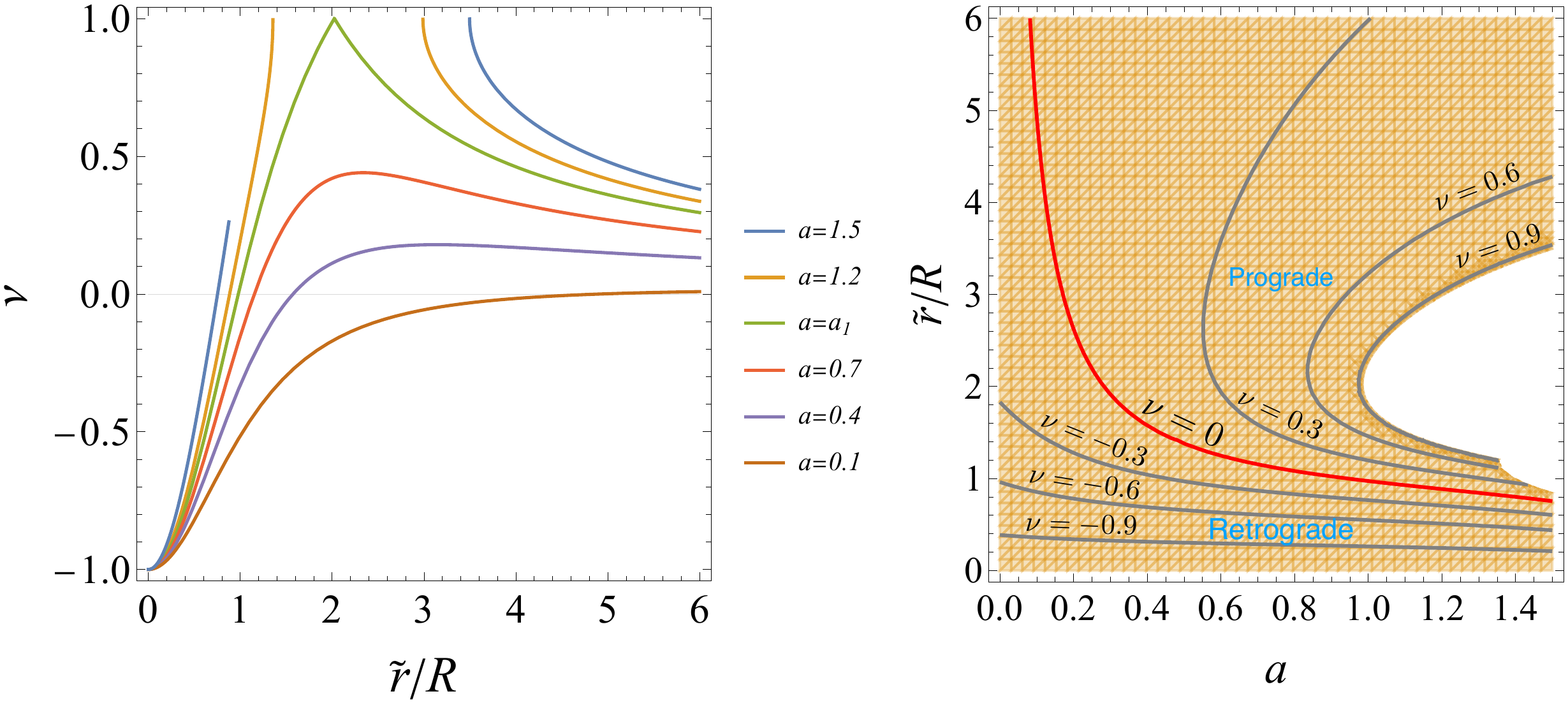}
\caption{
Precession rate $\nu$ on the matter distribution. 
The left figure shows $\nu$ as a function of $\tilde{r}$ for several values of $a$, where $\tilde{r}$ is the equilibrium radius of nearly circular bound orbits. 
The right figure shows the contour plot of $\nu$ in the $(a, \tilde{r}/R)$ plane. 
The red contour denotes $\nu=0$. The contour interval is $0.3$.}
\label{fig:nu}
\end{figure}

Figure~\ref{fig:nu} shows the behavior of $\nu$. 
The left figure shows $\nu$ as a function of the equilibrium radius $\tilde{r}$ for several values of $a$, and the right figure shows the contour plot of $\nu$ in the $(a, \tilde{r}/R)$ plane.
In the region sufficiently far from the center (i.e., $\tilde{r}\gg R$), the region $\nu>0$ appears, where the prograde periapsis shifts occur. 
This implies that the general-relativistic effect dominates over the local-density effect in the asymptotic region. 
In contrast, the region $\nu<0$ appears near the center (i.e., $\tilde{r}\ll R$). 
This implies that the local-density effect dominates over the general-relativistic effect near the center. 
This behavior of $\nu$ is independent of the value of $a$.
Note that the region $\nu<0$ extends to $\tilde{r}>R$ in the case of $a\ll1$. 
This is consistent with the consideration in the previous section. 
Furthermore, we can see that the function $\nu$ converges to $-1$ in the limit $\tilde{r}\to 0$ regardless of the value of $a$.
In this limit, the frequencies $\omega_\varphi$ and $\omega_r$ behave as $\omega_\varphi \to \omega_0$ and $\omega_r\to 2\omega_0$, respectively, where 
\begin{align}
\label{eq:omega0}
\omega_{0}=\frac{8}{(2+a)^2}\sqrt{\frac{a k}{4-a^2}}.
\end{align}
This result implies that a star moves in a two-dimensional isotropic harmonic oscillator potential near the center.

\section{Summary and discussion}
\label{sec:5}
We have considered the dynamics of a freely falling star
in the Buchdahl spacetime. 
To prevent anomalous effects due to matter distributions from affecting the dynamics, we have restricted the parameter range where all the energy conditions are satisfied in the entire spacetime region. 
Except for the system scale, the geometry is characterized by one parameter, the compactness, given by the ratio of the total mass to the scale of the dense core region. 
For small compactness, stable circular orbits of stars can exist in the entire region, whereas for large compactness, unstable circular orbits and the absence of circular orbits occur near the center, which are the signs of the general-relativistic effect.
Furthermore, 
the stable circular orbits exist near the center independent of the compactness 
because the stellar dynamics is generally reduced to the two-dimensional isotropic harmonic oscillation around the regular center.

We have particularly focused on how the competing effects of the local matter density and general relativity determine the periapsis shifts of nearly circular bound orbits. 
The asymptotic analysis of the effective potential allows us to identify two competing effects on the bounded motion: the general-relativistic effect and a local-density effect of matter. Furthermore, in the quasi-Newtonian analysis, we have explicitly demonstrated that the precession rate 
is determined by positive contribution of the general-relativistic effect and negative contribution of the local-density effect of matter. 
In fact, the analysis in the full geometry shows that 
the prograde periapsis shifts occur in the region sufficiently far from the center, where 
the general-relativistic effect dominates over the local-density effect. 
In contrast, the retrograde periapsis shifts occur near the center, where the local-density effect dominates over the general-relativistic effect. In particular, 
for small compactness, the retrograde periapsis shifts are observed even in the region outside the core radius.
Recently, the retrograde periapsis shift was found near the center of a numerical spacetime solution modeling dark matter distribution~\cite{Arguelles:2021jtk}, which is qualitatively consistent with our result.

If the observation of the shift angle $\Delta \varphi_{\mathrm{p}}$ for S2/S0-2 near Sgr A$^\ast$
is consistent with general relativity~\cite{Abuter:2020dou}, then it must satisfy at least $\Delta \varphi_{\mathrm{p}}>0$, or equivalently, $\nu>0$. This inequality leads to an upper bound of the core radius of the dark matter distribution,
\begin{align}
R < \sqrt{2M \tilde{r}}\, \approx \, 3 \mathrm{\, au} \left(\frac{M}{4\times 10^{6} M_{\odot}}
\right)^{1/2}\left(\frac{\tilde{r}}{120 \mathrm{\, au}}\right)^{1/2}. 
\end{align}
The upper limit is still $1$ order of magnitude larger than the gravitational radius of Sgr A$^\ast$ (i.e., $3 \mathrm{\, au}\approx 77M >2M$). This result still allows the possibility that the central object is a dark matter core.

Recently, a bright spot has been observed orbiting in the vicinity of Sgr A$^\ast$~\cite{GRAVITY:2018ofz}. Assuming that the motion of this spot is geodesic circular motion, we verify the existence of model parameters that reproduce this phenomenon. 
If we choose $a=0.55$ and $R=0.1 \mathrm{\, au}$, then we obtain 
the orbital period $T\approx 60 \mathrm{\, min}$, 
the orbital radius $\tilde{r}\approx0.4 \mathrm{\, au}\approx 8 \mathrm{\,kpc} \times 50 \mathrm{\, \mu as}$, 
and the total mass $M\approx4\times 10^6 M_\odot$. 
Furthermore, we confirm whether these model parameters consistently explain the observational results of S2/S0-2. Evaluating the proper mass $m$ at the periapsis and apoapsis distances of S2/S0-2, we have $m(120\mathrm{\, au})\approx m(1900\mathrm{\, au})\approx M$. Then, the mass fraction is $m(1900\mathrm{\, au})/m(120\mathrm{\, au})-1=O(10^{-7})\ll 1\%$. 
These results suggest that this dark matter model with a dense core is a possible alternative to a black hole that explains the observations of S2/S0-2 and the hot spot consistently.

Recently, several models of a black hole surrounded by matter distribution have been constructed in the class of the Einstein cluster~\cite{Cardoso:2021wlq,Jusufi:2022jxu,Igata:2022rcm}.
Moreover, periapsis shifts of stellar motion due to the local-density effect were studied in this system~\cite{Igata:2022rcm}, and its feature is similar to the present case. 
Our study implies that we need more information than periapsis shifts to 
distinguish a black hole surrounded by matter distribution from a dark matter distribution with a dense core.

\begin{acknowledgments}
The authors are grateful to Tomohiro Harada, Hideki Ishihara, Satoshi Iso, Kazunori Kohri, Takahiko Matsubara, Kouji Nakamura, Ken-ichi Nakao, Hiromi Saida, and Chul-Moon Yoo for useful comments and discussion. 
This work was supported by JSPS KAKENHI Grants No.~JP19K14715 and 
No.~JP22K03611 (T.~I.) and No. JP19H01900 and No. JP19H00695 (Y.~T.).
\end{acknowledgments}

\end{document}